\documentclass[12pt,preprint]{aastex}
\usepackage{graphics}
\usepackage{psfig}
\usepackage{epsfig}
\shorttitle{Survival of dark matter halos}
\shortauthors{Natarajan et al.}
\def\be{\begin{equation}}
\def\ee{\end{equation}}

\def\msun{{M_\odot}}

\def\gsim{\lower.5ex\hbox{\gtsima}}
\def\lsim{\lower.5ex\hbox{\ltsima}}
\def\gtsima{$\; \buildrel > \over \sim \;$}
\def\ltsima{$\; \buildrel < \over \sim \;$}
\def\prosima{$\; \buildrel \propto \over \sim \;$}
\def\gsim{\lower.5ex\hbox{\gtsima}}
\def\lsim{\lower.5ex\hbox{\ltsima}}
\def\simgt{\lower.5ex\hbox{\gtsima}}
\def\simlt{\lower.5ex\hbox{\ltsima}}
\def\simpr{\lower.5ex\hbox{\prosima}}

\def\beq#1{\begin{equation}\label{#1}}
\def\eeq{\end{equation}}
\def\bea#1{\begin{eqnarray}\label{#1}}
\def\bea{\begin{eqnarray}}
\def\eea{\end{eqnarray}}

\def\H2p{H$_2^+$ }

\def\mH2p{H_2^+}

\input psfig
\begin{document}
\title{The survival of dark matter halos in the cluster Cl\,0024+16}
\author{Priyamvada Natarajan, \altaffilmark{1,2,3}  Jean-Paul Kneib, \altaffilmark{4}  
Ian Smail, \altaffilmark{5} Tommaso Treu, \altaffilmark{6} 
 Richard Ellis, \altaffilmark{7,8}  
 Sean Moran, \altaffilmark{9} Marceau 
Limousin \altaffilmark{10} \& Oliver Czoske \altaffilmark{11}}

\affil{\footnotesize $^1$ Radcliffe Institute for Advanced Study, Harvard University,
10 Garden Street, Cambridge, MA 02138, USA}
\affil{\footnotesize $^2$ Department of Astronomy, Yale University, P.O. Box 208101, New
Haven, CT 06520-8101, USA} 
\affil{\footnotesize $^3$ Department of Physics, Yale University,
P.O. Box 208120, New Haven, CT 06520-8120, USA} 
\affil{\footnotesize $^4$ Observatoire Astronomique De Marseille Provence, Marseille, France}
\affil{\footnotesize $^5$ Institute for Computational Cosmology,
Durham University, South Road, Durham DH1 3LE, UK}
\affil{\footnotesize $^6$ Department of Physics, University of
California, Santa Barbara, CA 93106-9530, USA} 
\affil{\footnotesize $^7$ Department of Astronomy, Caltech, 1200 East California Blvd.,
Pasadena, CA 91125, USA} 
\affil{\footnotesize $^8$ Department of Astrophysics, Oxford University, Keble Road, 
Oxford OX1 3RH, UK}
\affil{\footnotesize $^9$ Department of Physics \& Astronomy, Johns Hopkins University,
366 Bloomberg Center, 3400 N. Charles Street, Baltimore, MD 21218, USA}
\affil{\footnotesize $^{10}$ Laboratoire d'Astrophysique de Toulouse et Tarbes, Tarbes,
France}
\affil{\footnotesize $^{11}$ Kapteyn Astronomical Institute, P. O. Box 800, 9700 AV 
Groningen, The Netherlands}

\begin{abstract}
  Theories of structure formation in a cold dark matter dominated
  Universe predict that massive clusters of galaxies assemble from the
  hierarchical merging of lower mass subhalos.  Exploiting strong and
  weak gravitational lensing signals inferred from panoramic {\it
  Hubble Space Telescope} imaging data, we present a high resolution
  reconstruction of the mass distribution in the massive, lensing
  cluster Cl\,0024+16 at $z = 0.39$. Applying galaxy-galaxy lensing
  techniques we track the fate of dark matter subhalos as a function
  of projected cluster-centric radius out to 5 Mpc, well beyond the virial
  radius. We report the first detection of the statistical lensing
  signal of dark matter subhalos associated with late-type galaxies in
  clusters. The mass of a fiducial dark matter halo that hosts an
  early type L$^*$ galaxy varies from $M =
  6.3_{-2.0}^{+2.7}\,\times\,10^{11}\,M_{\odot}$ within $r < 0.6$ Mpc,
  $1.3_{-0.6}^{+0.8}\,\times\,10^{12}\,M_{\odot}$ within $r < 2.9$ Mpc
  and increases further to $M =
  3.7_{-1.1}^{+1.4}\,\times\,10^{12}\,M_{\odot}$ in the outskirts. The
  mass of a typical dark matter subhalo that hosts an $L^*$ galaxy
  increases with projected cluster-centric radius in line with expectations from
  the tidal stripping hypothesis.  The mass of a dark matter subhalo
  that hosts a late-type L$^*$ galaxy is
  $1.06_{-0.41}^{+0.52}\,\times\,10^{12}\,M_{\odot}$. Early-type
  galaxies appear to be hosted on average in more massive dark matter
  subhalos compared to late-type galaxies. Early-type galaxies also
  trace the overall mass distribution of the cluster whereas late-type
  galaxies are biased tracers. We interpret our findings as evidence
  for the active assembly of mass via tidal stripping in galaxy
  clusters. The mass function of dark matter subhalos as a function of
  projected cluster-centric radius, is compared with an equivalent mass function
  derived from clusters in the Millenium Run simulation populated with
  galaxies using semi-analytic models. The shape of the
  observationally determined mass functions based on an I-band
  selected sample of cluster members and lensing data are in 
  agreement with the shapes of the subhalo mass functions derived from
  the Millenium Run simulation. However, simulated subhalos appear to
  be more efficiently stripped than lensing observations suggest. This
  is likely an artifact of comparison with a dark matter only
  simulation. Future simulations that simultaneously follow the
  detailed evolution of the baryonic component during cluster assembly
  will be needed for a more detailed comparison.
\end{abstract}
%%%%%%%%%%%%%%%%%%%%%%%%%%%%%%%%%%%%%%%%%%%%%%%%%%%

\section{Introduction}

Clusters of galaxies are the most massive and recently assembled
structures in the Universe. In the context of the hierarchical growth
of structure in a cold dark matter dominated Universe, clusters are
the repository of copious amounts of the dark matter. Gravitational
lensing, predicted by Einstein's theory of General Relativity, is the
deflection of light rays from distant sources by foreground mass
structures. In its most dramatic manifestation, strong lensing
requires a rare alignment with foreground dense structures and
produces highly distorted, magnified and multiple images of a single
background source (Schneider, Ehlers \& Falco 1992). More commonly,
the observed shapes of background sources viewed via a foreground
cluster lens are systematically elongated, in the so-called weak
lensing regime. Strong and weak lensing offer the most reliable probes
of the distribution of dark matter on various cosmic scales (Blandford
\& Narayan 1992; Mellier 2002; Schneider, Ehlers \& Falco 1992).

Strong lensing studies of the core regions of several clusters
indicate that the dark matter distribution can be represented by a
combination of smoothly distributed, extended cluster mass components and
smaller-scale clumps or subhalos associated with luminous galaxies
(Kneib et al. 1996; Natarajan \& Kneib 1997; Natarajan, Kneib, Smail
\& Ellis 1998). The smooth components have been detected using weak
lensing techniques out to the turn-around radius (typically of the
order of several Mpc) in clusters (Kneib et al. 2003; Gavazzi et al. 2003; 
Broadhurst et al. 2005; Bradac et al. 2006; Clowe et al. 2006; 
Wittman et al. 2006; Limousin et al. 2007c; Bardeau et al. 2007). To date,
however, attention has largely focused on the lensing derived density
profile of the smooth cluster component, and its agreement with profiles
computed from high resolution numerical simulations of structure
formation in the Universe (Navarro, Frenk \& White 1997; Navarro et
al. 2004; Sand et al. 2004). In fact, the granularity of the dark
matter distribution associated with individual galactic subhalos holds
important clues to the growth and assembly of clusters. Several
earlier studies have explored this issue for the particular case of
Cl\,0024+16 (Tyson, Kochanski \& dell'Antonio 1998; Broadhurst, Huang,
Frye \& Ellis 2000; Jee et al. 2007; Smail et al. 1996).

The detailed mass distribution of clusters and in particular, the
fraction of the total cluster mass associated with individual galaxies
has important consequences for the frequency and nature of galaxy
interactions in clusters (Merritt 1983; Moore et al.\ 1996; Ghigna et
al.\ 1998; Okamato \& Habe 1999). Infalling subhalos suffer a range of
violent fates as the strong gravitational potential of the cluster
tidally strips dark matter and removes baryons via ram-pressure
stripping from them (Cortese et al. 2007). Simulations suggest that
subhalos may not be arranged equally around galaxies of different
morphologies given their varying histories in the cluster environment
(Ghigna et al. 1998; Tormen, Diaferio \& Syer 1998; Springel, White,
Tormen \& Kauffmann 2001). Moreover, subhalos may become tidally
truncated by an amount that will differ substantially over the large
dynamic range in cluster density. Observations of tidal stripping
offer important clues to key questions regarding the growth and
evolution of clusters. How much dark matter is associated with the
subhalos in clusters as a function of radius?  To what extent do the
luminous cluster galaxy populations trace the detailed mass
distribution? And, how significant is tidal stripping for the various
morphological galaxy types in the cluster? These are questions we
attempt to answer in this work using observational data and by comparing
with numerical simulations. 

To explore cluster galaxy masses, we exploit the technique of
galaxy-galaxy lensing, which was originally proposed as a method to
constrain the masses and spatial extents of field galaxies (Brainerd,
Blandford \& Smail 1996), which has been since extended and developed
over the years to apply inside clusters (Natarajan \& Kneib 1996;
Geiger \& Schneider 1998; Natarajan et al. 1998; 2002a; Natarajan, De
Lucia \& Springel 2007; Limousin et al. 2007a).  Previous attempts to
measure the granularity of the dark matter distribution as a function
of cluster-centric radius from observations have had limited
success. Analyzing CFHT (Canada-France-Hawaii Telescope) weak lensing
data from the supercluster MS0302+17, Gavazzi et al.\ (2004) claimed
detection of a radial trend in the extents of dark matter subhalos in
this supercluster region extending out to a few Mpc from the
center. Gavazzi et al. reported that the mass distribution derived
from weak lensing was robustly traced by the luminosity of early-type
galaxies, although their analysis did not include late-type galaxies
or a large-scale smooth component. However, utilizing ground-based
CFHT weak lensing data for a sample of massive clusters at $z = 0.2$,
Limousin et al.\ (2005) did not detect any variation of the dark
matter subhalo masses with cluster-centric radius out to a significant
fraction of the virial radius. The resolution of ground-based data
appears to be inadequate to detect this effect.

In this paper, we present the determination of the mass function of
substructure in Cl\,0024+16 (at $z = 0.39$) in 3 radial bins using
panoramic Hubble Space Telescope {\it HST} imaging data. A high
resolution mass model tightly constrained by current observations is
constructed including individual cluster galaxies and their associated
dark matter subhalos. We show that over a limited mass range we can
successfully construct the mass function of subhalos inside this
cluster as a function of cluster-centric radius. The three bins span
from the center to 5 Mpc (well beyond the the virial radius of 1.7
Mpc) providing us insights into the tidal stripping process. We also
compare properties of the subhalos that host early-type galaxies with
those that host late-type galaxies in Cl\,0024+16. In addition, we
compare the results retrieved from the lensing analysis with results
from the largest cosmological simulation carried out so far - the {\it
Millennium Simulation}. N-body simulations in combination with
the semi-analytic models that we employ in this work are an invaluable
tool for investigating the non-linear growth of structure in detail
and to provide insights into the cluster assembly process.

The outline of this paper is as follows: in \S2, we discuss the
theoretical framework of tidal stripping and galaxy-galaxy lensing in
clusters; in \S3 the observations and modeling are described. The
analysis for Cl\,0024+16 is presented in \S4 including a discussion of
the uncertainties; results and the comparison with clusters in the
Millennium Simulation are described in \S5. We conclude with a
discussion of the implications of our results for the LCDM model and
the future prospects of this work.
\footnote{Throughout this work wherever required we have used the
following values for the cosmological parameters: H$_0$ = 72 km/s/Mpc;
$\Omega_{\rm m}$ = 0.3; $\Omega_{\Lambda}$ = 0.7.  At the redshift of
Cl\,0024+16, 1'' = 5.184 kpc.} 

\section{Theoretical framework}

\subsection{Tidal stripping and dynamical modification in clusters}

Theoretical studies of cluster formation using simulations and
analytic models predict that there are two key dynamical processes
(Ghigna et al. 1998; Springel, White, Tormen \& Kauffmann 2001; De
Lucia et al. 2004; Moore, Katz, Lake, Dressler \& Oemler 1996; Balogh,
Navarro \& Morris 2000; Merritt 1985) that are relevant to the mass
loss of infalling dark matter subhalos in assembling clusters. The
first process is tidal stripping induced by the interaction of
infalling galaxies and groups with the global tidal field generated by
the smooth dark matter distribution. The second process is
modification to the mass distribution due to high and low velocity
encounters between infalling subhalos (Moore, Katz, Lake, Dressler \&
Oemler 1996).

For the purposes of studying the dynamics of galaxies in clusters we have
partitioned the cluster into 3 distinct regions: the inner
core region where the global tidal field is the strongest, the
transition region where the two above mentioned dynamically transformative
processes occur and finally, the periphery where the dominant
stripping is due to interactions between the infalling galaxies and
groups rather than the global tidal field (Treu et al. 2003). Detailed
study of the properties of cluster galaxies in Cl\,0024+16 by Treu et al. 2003,
find that demarcation into these 3 regions is naturally provided by the 
dynamical processes that operate efficiently at various radii from the 
cluster center.

In the central region, the gravitational potential of the cluster is
the strongest and tidal stripping is expected to be the dominant
dynamically transformative process. Recent tidal effects are not
expected in the transition region whereas most galaxies inhabiting the
periphery are likely to have never traversed the cluster center.  The
galaxies in the outer regions are expected to be modified
predominantly due to local interactions with other nearby galaxies and
groups despite being gravitationally bound to the cluster.

An analytic estimate of the effect of tidal truncation as a function
of cluster-centric radius can be calculated by modeling Cl\,0024+16 as
an isothermal mass distribution and considering the motions of cluster
subhalos in this potential (Merritt 1985). In this framework the tidal
radius of a subhalo hosting a cluster galaxy is given by:
\begin{eqnarray}
R_{\rm tidal} \propto (\frac{\sigma_{\rm gal}}{\sigma_{\rm cluster}})\,r,
\end{eqnarray} 
where $R_{\rm tidal}$ is the tidal radius of the subhalo, $\sigma_{\rm
gal}$ is the central velocity dispersion of the galaxy, $\sigma_{\rm
cluster}$ is the velocity dispersion of the cluster and $r$ is the
distance from the cluster center. The current paradigm for structure
formation in the Universe predicts that the masses of infalling
subhalos are a strong function of cluster-centric radius $r$,
indicative of the variation of the strength of tidal stripping from
the periphery (where it is modest) to the inner regions, where it is
severe (Springel, White, Tormen \& Kauffmann 2001; De Lucia et
al. 2004; Moore, Katz, Lake, Dressler \& Oemler 1996; Balogh, Navarro
\& Morris 2000). Mapping the mass function of subhalos directly from
observations offers a powerful way to test these theoretical
predictions.

\subsection{Galaxy-galaxy lensing in clusters}

In this subsection we briefly outline the analysis framework. Details
can be found in several earlier papers (Natarajan \& Kneib 1996;
Natarajan et al. 1998; Natarajan et al. 2002a; 2002b; Natarajan, De
Lucia \& Springel 2007). For the purpose of constraining the
properties of the subhalo population, Cl\,0024+16 is modeled
parametrically as a super-position of smooth large-scale mass
components, which we will refer to with subscript `s' hereafter, and
smaller scale potentials that are associated with bright cluster
members, referred to as perturbers denoted by the subscript
`$p_i$'. Using the same data-set to construct a mass distribution for
Cl\,0024+16 Kneib et al. (2003) found that the best-fit model required
two large-scale components. In our current modeling, we adopt that
parametrization as the prior. In earlier work, our analysis was
limited by data to the inner regions of clusters ($<$ 1 Mpc), and only
to early-type galaxies as perturbers as a consequence (Natarajan et
al. 1998; Natarajan, Kneib \& Smail 2002; Natarajan, De Lucia \&
Springel 2007). With the current data-set we also probe the late-type
cluster member population and statistically constrain parameters that
characterize their dark matter subhalos. There are however, an
insufficent number of late-types in the core region, their numbers
steadily increase with cluster-centric radius. Therefore, in the core
region, we focus on the subhalos of early-types. In effect, the contribution
of late-types in the core region gets inevitably taken into account as
part of the smooth mass distributions. We note here that while we illustrate
our formalism with simple equations to provide insight into our framework, 
ultimately the analysis is performed numerically and all the non-linearities 
arising in the lensing inversion are taken into account. The gravitational
potential of Cl\,0024+16 is modeled as follows:
\begin{equation}
\phi_{\rm tot} = \Sigma_{n}\,\phi_{\rm s} + \Sigma_i \,\phi_{\rm p_i},
\end{equation}
where the two $\phi_{\rm s}$ ($n=1$ and $n=2$) components represent
the potentials that characterize the smooth component and $\phi_{\rm
p_i}$ are the potentials of the galaxy subhalos treated as
perturbers. The corresponding deflection angle $\alpha_I$ and the
amplification matrix $A^{-1}$ can also be decomposed into independent
contributions from the smooth clumps and perturbers,
\begin{eqnarray}
\alpha_I\,=\,\Sigma_n\,{{\mathbf \nabla}\phi_{\rm s}}\,+\,\Sigma_i \,
{{\mathbf \nabla}\phi_{\rm p_i}},\,\,\\ \nonumber
\,A^{-1}\,=\,I\,-\,\Sigma_n\,{{\mathbf 
\nabla\nabla} {\phi_{\rm s}}}\,-\,\Sigma_i \,{{\mathbf \nabla\nabla} 
{\phi_{\rm p_i}}}.
\end{eqnarray}
In fact, the amplification matrix can be decomposed as a linear
sum: 
\begin{eqnarray} 
A^{-1}\,=\,(1\,-\,\Sigma_n\,\kappa_{\rm s}\,-\,\Sigma_i \kappa_{\rm
p})\,I - \Sigma_n\,\gamma_{\rm s}J_{2\theta_{\rm s}} - \Sigma_i \,\gamma_{\rm
p_i}J_{2\theta_{\rm p_i}}, 
\end{eqnarray}
where $\kappa$ is the magnification
and $\gamma$ the shear.  The shear $\gamma$ is written as a complex
number and is used to define the reduced shear $\overline{g}$, 
which is the quantity that is measured directly from observations 
of the shapes of background galaxies. The reduced shear
can also be further decomposed into contributions from the smooth pieces
and the perturbers:
\bea
\overline{g_{tot}} = {\overline{\gamma} \over 1-\,\kappa} =
{{\Sigma_n\,\overline\gamma_{\rm s}} + \Sigma_i \,{\overline\gamma_{p_i}} \over
1-\Sigma_n\,\kappa_{\rm s} -\Sigma_i \,\kappa_{p_i}},
\eea 
Here $\bar{\gamma}$ is the mean shear of background galaxies in an annulus
around a particular early-type cluster galaxy treated as a local perturber.
In the frame of an individual perturber $j$ (neglecting effect 
of perturber $i$ if $i \neq j$), the above simplifies to: 
\bea 
{\overline g_{tot}}|_j} = { {\Sigma_n\,{\overline \gamma_{\rm s}}
+{\overline \gamma_{p_j}} \over {1-\Sigma_n\,\kappa_{\rm s} -\kappa_{p_j }}}. 
\eea
Restricting our analysis to the weak regime (as mentioned above the analysis
is ultimately performed numerically and includes the effect of strong lensing), 
and thereby retaining only the first order terms from the lensing equation for the shape
parameters (e.g. Kneib et al.\ 1996) we have: 
\be 
{\overline g_I}=
{\overline g_S}+{\overline g_{tot}}, 
\ee 
where ${\overline g_I}$ is the distortion 
of the image, ${\overline g_S}$ the intrinsic shape of the source, 
${\overline g_{\rm tot}}$
is the distortion induced by the lensing potentials (the smooth
component as well as the perturbers). Note that the equations are
outlined here to provide a feel for the technique. The lensing
inversion for the observational data is done numerically taking the
full non-linearities that rise in the strong lensing regime into
account. \footnote{The measured image shape and orientation are used
to construct a complex number whose magnitude is given in terms of the
semi-major axis (a) and semi-minor axis (b) of the image and the
orientation is the phase of the complex number.}

In the local frame of reference of the subhalos, the mean value of the
quantity ${\overline g_I}$ and its dispersion are computed in circular
annuli (at radius $r$ from the perturber centre), assuming a known 
value for the smooth cluster component over the area of integration. 
In the frame of the perturber, the averaging procedure allows efficient 
subtraction of the large-scale component, enabling the extraction of the 
shear component induced in the background galaxies only by the local 
perturber. The background galaxies are assumed to have intrinsic 
ellipticities drawn from a known distribution (see the Appendix for further
details). Schematically the effect of the cluster on the intrinsic
ellipticity distribution of background sources is to cause a coherent
displacement and the presence of perturbers merely adds small-scale
noise to the observed ellipticity distribution. Since we are subtracting
a long-range signal to statistically extract a smaller scale anisotropy
riding on it, we are inherently limited to physical scales on which the
contrast is maximal, i.e. galaxy subhalo scales. 

The contribution of the smooth cluster component has 2 effects: it
boosts the shear induced by the perturber which becomes non-negligible
in the cluster center, and it simultaneously dilutes the regular
galaxy-galaxy lensing signal due to the ${\sigma^2_{\overline g_{\rm
s}} / 2}$ term in the dispersion. However, one can in principle
optimize the noise by `subtracting' the measured cluster signal
${\overline g_{\rm s}}$ using a tightly constrained parametric model
for the cluster.

The feasibility of this differenced averaging prescription for
extracting the distortions induced by the possible presence of dark
matter subhalos around cluster galaxies with {\it HST} quality data
has been amply demonstrated in our earlier papers (Natarajan et al.\
1998; 2002a; 2004; 2007). We have also shown with direct comparison to
simulations that the we can reliably recover substructure mass
functions with this technique in the inner 1 Mpc or so of galaxy
clusters. Note here that it is the presence of the underlying
large-scale smooth mass components (with a high value of $\kappa_s$)
that enables the extraction of the weaker signal riding on it.
\section{Observations and modeling}

\subsection{The HST WFPC-2 data-set}

Our dataset comprises a mosaic of 39 sparsely-sampled images of the
rich cluster Cl\,0024+16 ($z = 0.39$) taken by the Wide Field
Planetary Camera-2 on the Hubble Space Telescope ({\it HST}). By applying
lensing techniques to this panoramic imaging dataset, we aim to
characterize the fine scale distribution of dark matter. This unique
dataset extends to the turn-around radius $\simeq$5 Mpc, well beyond
the inner 0.5 - 1 Mpc that has been studied previously. This enables
us to map the detailed dark matter distribution and to calibrate the
tidal stripping effect as a function of distance from the cluster
center. In earlier analysis, we combined strong and weak lensing
constraints to provide an accurate representation of the smooth dark
matter component out to 5 Mpc radius (Kneib et al. 2003). Strong
lensing provides stringent constraints on the mass profile in the
inner region while the detected weak shear constrains the profile out
at large radii (Mellier 2002; Kneib et al. 2003). Non-contiguous,
sparse sampling of the {\it HST} pointings was chosen to maximize radial
coverage. Further details of the data and analyses can be found in
earlier published papers (Kneib et al. 2003; Treu et al. 2003).
We re-iterate here that the WFPC-2 data-set used for this analysis 
has been presented already and is described in detailed in earlier works 
by our group, including the determination of shapes for the background 
galaxies (in Kneib et al. 2003); selection and confirmation of cluster 
membership and morphological classifications of cluster galaxies 
(in Treu et al. 2003). 

Here, we use galaxy-galaxy lensing to detect cluster galaxy subhalos
associated with early-type galaxies and late-type galaxies against the
background of smoothly distributed dark matter in three radial bins.
Using the extensive set of ground-based spectra (Czoske et al. 2001;
Czoske, Moore, Kneib \& Soucail 2002; Moran et al. 2005) and {\it HST}
morphologies (Treu et al. 2003), we first identified early-type and
late-type members to well beyond the virial radius, ($r_{\rm vir} =
1.7\,{\rm Mpc}$), out to $\sim\,5\,{\rm Mpc}$. Details of the data
reduction, cluster membership determination and morphological
classification can be found in Treu et al.(2003).

\subsection{Modeling the cluster Cl\,0024+16}

Cl\,0024+16 is an extremely massive cluster and has a surface mass
density in the inner regions which is significantly higher than the
critical value, therefore produces a number of multiple images of
background sources. By definition, the critical surface mass density
for strong lensing is given by:
\begin{eqnarray}
\Sigma_{\rm crit} = {\frac{c^2}{4 \pi G}} \frac{D_s}{D_d D_{ds}}
\end{eqnarray}
where $D_s$ is the angular diameter distance between the observer and the
source, $D_d$ the angular diameter distance between the observer and
the deflecting lens and $D_{ds}$ the angular diameter distance between
the deflector and the source. 

Note that the integrated lensing signal detected is due to all the
mass distributed along the line of sight in a cylinder projected onto
the lens plane. In this and all other cluster lensing work, the
assumption is made that individual clusters dominate the lensing
signal as the probability of encountering two massive rich clusters
along the same line-of-sight is extremely small due to the fact that
these are very rare objects in hierarchical structure formation
models.  Cl\,0024+16 is known to have a significant amount of
substructure in velocity space. Czoske et al. (2002) and more recently
Moran et al. (2005) have performed comprehensive redshift surveys of
this cluster and its environs and have enabled the construction of a
3-dimensional picture for this cluster using the $\sim 500$ galaxy
redshifts within about 3 - 5 Mpc from the cluster center. Their
combined data reveal a foreground component of galaxies separated from
the main cluster in velocity space. Both groups argue that this is
likely a remnant of a high-speed collision between the main cluster
and an infalling sub-cluster. The detailed redshift distribution of
cluster members in Cl\,0024+16 is taken carefully into account in our
lensing analysis, starting with a prior that includes 2 large-scale
components to model the smooth mass distribution.

With our current sensitivity limits, galaxy-galaxy lensing within the
cluster provides a determination of the total enclosed mass within an
aperture. We lack sufficient sensitivity to constrain the detailed
mass profile for individual cluster galaxies. With higher resolution
data in the future we hope to be able to obtain constraints on the
slopes of mass profiles within subhalos. In this paper, the subhalos
are modeled as pseudo-isothermal elliptical components (PIEMD models,
derived by Kassiola \& Kovner 1993) centered on galaxies that lie
within a projected radius of out to 5 Mpc from the cluster center and
two NFW profiles are used to model the smooth, large-scale
contribution.  We find that the final results obtained for the
characteristics of the subhalos (or perturbers) is largely independent
of the form of the mass distribution used to model the smooth,
large-scale components. A comparison of the best-fit profiles for the
smooth component from lensing with those obtained in high resolution
cosmological N-body simulations has been presented in the work of
Kneib et al. (2003). Combining strong and weak constraints, they were
able to probe the mass profile of the cluster on scales of 0.1-5 Mpc,
thus providing a valuable test of the universal form proposed by
Navarro, Frenk, \& White (NFW) on large scales. We use the best-fit
mass model of Kneib et al. (2003) for the smooth component as a prior
in our analysis, although we allow the parameters like the centroids
of the 2 large-scale components and their velocity dispersion to vary
when obtaining constraints on the subhalos. The 2 NFW components used
as priors are characterized by the following properties: Clump 1: with
$M_{200} = 6.5 \times 10^{14}\,\msun$; $c = 22_{-5}^{+9} $; $r_{200} =
1.9\,{\rm Mpc} $; $r_s = 88\,{\rm kpc}$ and Clump 2: with $M_{200} =
2.8 \times 10^{14}\,\msun$; $c = 4_{-1}^{+2}$; $r_{200} = 1.5\,{\rm
Mpc}$; $r_s = 364\,{\rm kpc} $.

To quantify the lensing distortion induced, the individual
galaxy-scale halos are modeled using the PIEMD profile with,
\begin{eqnarray}
\Sigma(R)\,=\,{\Sigma_0 r_0  \over {1 - r_0/r_t}}
({1 \over \sqrt{r_0^2+R^2}}\,-\,{1 \over \sqrt{r_t^2+R^2}}),
\end{eqnarray}
with a model core-radius $r_0$ and a truncation radius $r_t\,\gg\,
r_0$. Correlating the above mass profile with a typical de
Vaucleours light profile (the observed profile for bright early type
galaxies) provides a simple relation between the truncation radius and
the effective radius $R_{\rm e}$, $r_t\sim (4/3) R_{\rm e}$. The
coordinate $R$ is a function of $x$, $y$ and the ellipticity, 
\bea
R^2\,=\,({x^2 \over (1+\epsilon)^2}\,+\,{y^2 \over (1-\epsilon)^2})\,;
\ \ \epsilon= {a-b \over a+b}, 
\eea 
The mass enclosed within an aperture radius $R$ for the $\epsilon = 0$ 
model is given by: 
\be 
M(R)={2\pi\Sigma_0
r_0 \over {1-{{r_0} \over {r_t}}}}
[\,\sqrt{r_0^2+R^2}\,-\,\sqrt{r_t^2+R^2}\,+\,(r_t-r_0)\,].  
\ee 
The total mass $M$, is finite with $M\,
\propto \,{\Sigma_0} {r_0} {r_t}$. 
The shear is:
\bea
\gamma(R)\,&=&\,\nonumber \kappa_0[\,-{1 \over \sqrt{R^2 + r_0^2}}\, +\,{2 \over
R^2}(\sqrt{R^2 + r_0^2}-r_0)\,\\ \nonumber &+&\,{1 \over {\sqrt{R^2 +
r_t^2}}}\,-\, {2 \over R^2}(\sqrt{R^2 + r_t^2} - r_t)\,].\\ 
\eea
In order to relate the light distribution in cluster galaxies to key
parameters of the mass model of subhalos, we adopt a set of physically
motivated scaling laws derived from observations (Brainerd et al.\
1996; Limousin et al. 2005; Halkola et al. 2007): 
\begin{eqnarray}
{\sigma_0}\,=\,{\sigma_{0*}}({L \over L^*})^{1 \over 4};\,\,
{r_0}\,=\,{r_{0*}}{({L \over L^*}) ^{1 \over 2}};\,\,
{r_t}\,=\,{r_{t*}}{({L \over L^*})^{\alpha}}.
\end{eqnarray}
The total mass $M$ enclosed within an aperture $r_{t*}$ and
the total mass-to-light ratio $M/L$ 
then scale with the luminosity as follows for the early-type galaxies:
\begin{eqnarray}
M_{\rm ap}\,\propto\,{\sigma_{0*}^2}{r_{t*}}\,({L \over L^*})^{{1 \over
2}+\alpha},\,\,{M/L}\,\propto\,
{\sigma_{0*}^2}\,{r_{t*}}\left( {L \over L^*} \right )^{\alpha-1/2},
\end{eqnarray}
where $\alpha$ tunes the size of the galaxy halo. In this work
$\alpha$ is taken to be $1/2$. These scaling laws are empirically
motivated by the Faber-Jackson relation for early-type galaxies
(Brainerd, Blandford \& Smail 1996). For late-type cluster members, we
use the analogous Tully-Fisher relation to obtain scalings of
$\sigma_{0*}$ and ${r_{t*}}$ with luminosity. The empirical
Tully-Fisher relation has significantly higher scatter than the
Faber-Jackson relation. In this analysis we do not take the scatter
into account while employing these scaling relations. We assume these
scaling relations and recognize that this could ultimately be a
limitation but the evidence at hand supports the fact that mass traces
light efficiently both on cluster scales (Kneib et al. 2003) and on
galaxy scales (McKay et al. 2001; Wilson et al. 2001; Mandelbaum et
al. 2006). Further explorations of these scaling relations have
recently been presented in Halkola \& Seitz (2007) and Limousin,
Sommer-Larsen, Natarajan \& Milvang-Jensen (2007). The redshift
distribution and intrinsic ellipticity distribution assumed for this
analysis are outlined in the Appendix.

\subsection{The maximum-likelihood method}

Parameters that characterize both the global components and the
subhalos are optimized, using the observed strong lensing features -
positions, magnitudes, geometry of multiple images and measured
spectroscopic redshifts, along with the smoothed shear field as
constraints. With the parameterization presented above, we optimize
and extract values for the central velocity dispersion and the
aperture scale $(\sigma_{0*}, r_{t*})$ for a subhalo hosting a
fiducial $L^*$ cluster galaxy.

Maximum-likelihood analysis is used to obtain significance bounds on
these fiducial parameters that characterize a typical $L^*$ subhalo in the
cluster. The likelihood function of the estimated probability
distribution of the source ellipticities is maximized for a set of
model parameters, given a functional form of the intrinsic ellipticity
distribution measured for faint galaxies. For each `faint' galaxy
$j$, with measured shape $\tau_{\rm obs}$, the intrinsic shape
$\tau_{S_j}$ is estimated in the weak regime by subtracting the
lensing distortion induced by the smooth cluster models and the galaxy
subhalos,
\begin{eqnarray}
\tau_{S_j} \,=\,\tau_{\rm obs_j}\,-{\Sigma_i^{N_c}}\,
{\gamma_{p_i}}\,-\, \Sigma_n\,\gamma_{c}, 
\end{eqnarray}
where $\Sigma_{i}^{N_{c}}\,{\gamma_{p_i}}$ is the sum of the shear
contribution at a given position $j$ from $N_{c}$ perturbers. This
entire inversion procedure is performed numerically using the code
developed that builds on the ray-tracing routine {\sc lenstool}
written by Kneib (1993). This machinery accurately takes into account
the non-linearities arising in the strong lensing regimeas well. Using a
well-determined `strong lensing' model for the inner-regions along
with the shear field and assuming a known functional form for
$p(\tau_{S})$ the probability distribution for the intrinsic shape
distribution of galaxies in the field, the likelihood for a guessed
model is given by,
\begin{eqnarray}
 {\cal L}({{\sigma_{0*}}},{r_{t*}}) = 
\Pi_j^{N_{gal}} p(\tau_{S_j}),
\end{eqnarray}
where the marginalization is done over $(\sigma_{0*},r_{t*})$.
We compute ${\cal L}$ assigning the median redshift corresponding to the
observed source magnitude for each arclet. The best fitting model 
parameters are then obtained by maximizing the log-likelihood 
function $l$ with respect to the parameters ${\sigma_{0*}}$ and ${r_{t*}}$. 
Note that the parameters that characterize the smooth component are
also simultaneously optimized. In this work, we perform this
likelihood analysis in each of the 3 radial bins to obtain the set of
$(\sigma_{0*},r_{t*})$ that characterize subhalos in each radial bin. 

In summary, the basic steps of our analysis therefore involve lens
inversion, modeling and optimization, which are done using the {\sc
lenstool} software utilities (Kneib 1993; Jullo et al. 2007). These
utilities are used to perform the ray tracing from the image plane to
the source plane with a specified intervening lens. This is achieved
by solving the lens equation iteratively, taking into account the
observed strong lensing features, positions, geometry and magnitudes
of the multiple images. In this case, we also include a constraint on
the location of the critical line (between 2 mirror multiple images)
to tighten the optimization. We fix the core radius of an $L^*$
subhalo to be $0.1\,{\rm kpc}$, as by construction our analysis cannot
constrain this quantity. The measured shear field and the measured
velocity dispersions of early-type galaxies are used as priors in the
likelihood estimation. In addition to the likelihood contours, the
reduced $\chi^2$ for the best-fit model is also found to be robust.

\section{Analysis for Cl\,0024+16}

To detect cluster subhalos, we first select a population of background
galaxies within a magnitude range $23\,<\,I\,<\,26$ (measured in the
F814W filter) and determined their individual shapes to a high degree
of accuracy taking into account the known anisotropy of the point
spread function of the WFPC-2 Camera (Bridle et al. 2002; Kuijken
1999). Details of this procedure and the systematics are described in
detail in Kneib et al. (2003). Shape distortions in this population
were then used to compute the masses of the foreground cluster and its
member subhalos.

To quantify environmental effects on infalling dark matter halos and
noting the three physical regimes discussed earlier, we divided the
cluster into three regions: the central region extending out to $\sim$
600 kpc from the center [core]; the transition region extending out to
$\sim 1.7\,r_{\rm vir} \sim 2.9\,{\rm Mpc}$, [transition] and the
periphery out to $\sim 2.8\,r_{\rm vir} \sim 4.8\,{\rm Mpc}$
[periphery]. These bins partition the cluster into regions of high,
medium and low galaxy number density and dark matter density (Treu et
al. 2003) respectively. The typical surface density of cluster members
in the core region is 120 galaxies per Mpc$^{-2}$; in the transition
region it drops to about 60 galaxies per Mpc$^{-2}$ and in the
periphery it is roughly 50 galaxies per Mpc$^{-2}$ (Treu et al. 2003).

\begin{figure*}
\centerline{\psfig{file=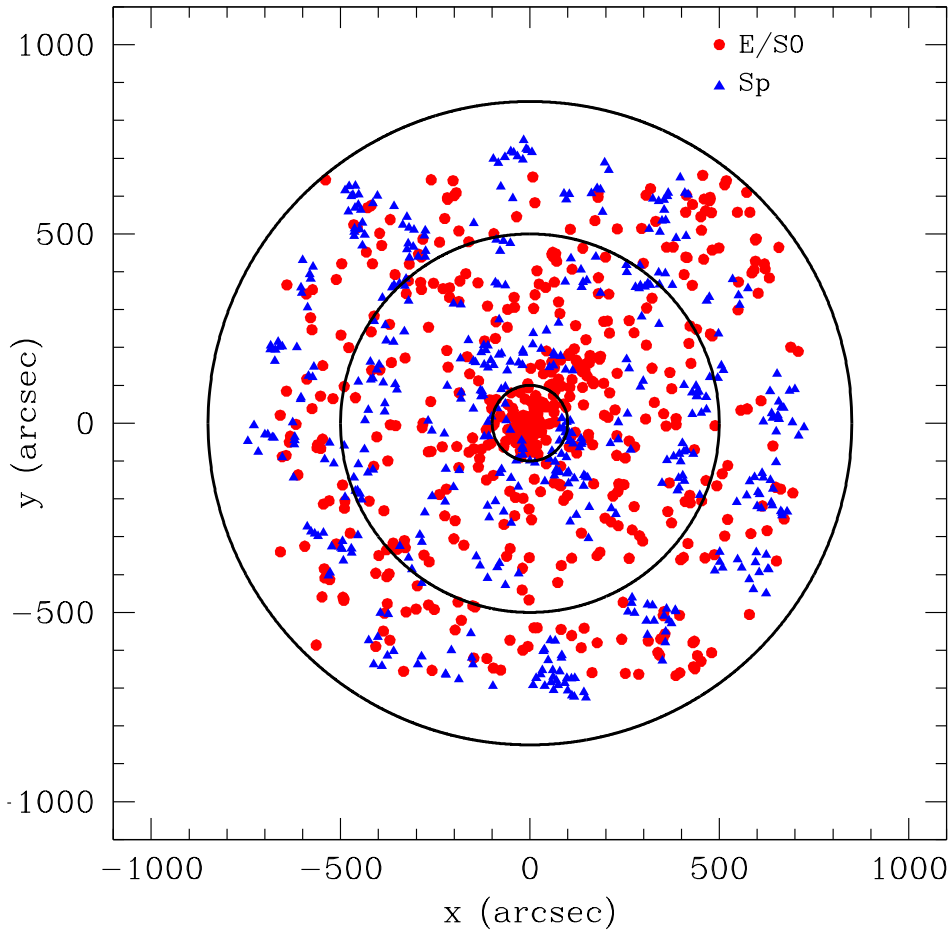,width=0.8\textwidth}}
\caption{The spatial distribution of morphologically classified
early-type galaxies (red circles) and late-type galaxies (blue
triangles) with measured redshifts in Cl\,0024+16 derived from the
sparsely sampled mosaic using the {\it WFPC-2} Camera aboard the {\it
HST}. The three circles define the radial binning used in our
analysis. The inner-most circle encompasses the {\bf core} region of
the cluster out to 0.6 Mpc, the middle circle the {\bf transition}
region extending out to 2.9 Mpc and the outer circle marks the {\bf
periphery} of the cluster out to 4.8 Mpc. Galaxies plotted here
include spectroscopically confirmed cluster members and galaxies with
secure photometric redshifts in the {\it HST} footprint.}
\end{figure*}

A well defined morphology-density relation is detected in Cl\,0024+16
(Treu et al. 2003; Dressler 1980; Fasano et al. 2000).  The fraction
of early-type galaxies declines steeply away from the center, starting
at 70-80\% out to 1 Mpc and decreasing down to 50\% at the
outskirts. In contrast, the late-type galaxy population fraction is
negligible in the center but increases in the transition region and
constitutes 50\% out at the periphery. In fact, Moran et
al.  (2007) find that the spirals are kinematically disturbed even
well beyond the virial radius in this cluster. In the core, cluster
membership was defined strictly, and only spectroscopically confirmed
members were used in the galaxy-galaxy lensing analysis. In the
transition region and the periphery, the classification of cluster
members was performed using both spectroscopically and photometrically
determined redshifts. We selected cluster galaxies within
$17\,<\,I\,<\,22$ to ensure comparable degrees of completeness for
both morphological types across all three bins. Our selection
procedure yields 51 early-types in the core; 93 in the transition
region [70 spectroscopically confirmed] and 44 [15 spectroscopically
confirmed] in the periphery. Including early-types from the ground
based survey work (Moran et al. 2007) we have an additional 257
members in the transition region and 294 members in the
periphery. There are a total of 331 late-types (this inventory
includes the {\it HST} mosaic and ground based data) confined to
transition and periphery region. For the early-types in the {\it HST WFPC-2}
mosaic, all 51 in the core are spectroscopically confirmed to be
cluster members, in the outer 2 bins, about 63$\pm$7\% of the
early-types are spectroscopically confirmed, and across all morphologies
$\sim$ 65\% have secure measured redshifts. In addition, we have
redshifts for early-type candidates that lie outside our tiled {\it HST}
mosaic as well photometric redshifts estimates.  The radial
distribution of the selected cluster galaxies is shown in
Figure~1. The similarity of the luminosity function of the selected
early-types in the three bins shown in Figure~2 ensures that we have
truly comparable samples with no luminosity bias.

\begin{figure*}
\centerline{\psfig{file=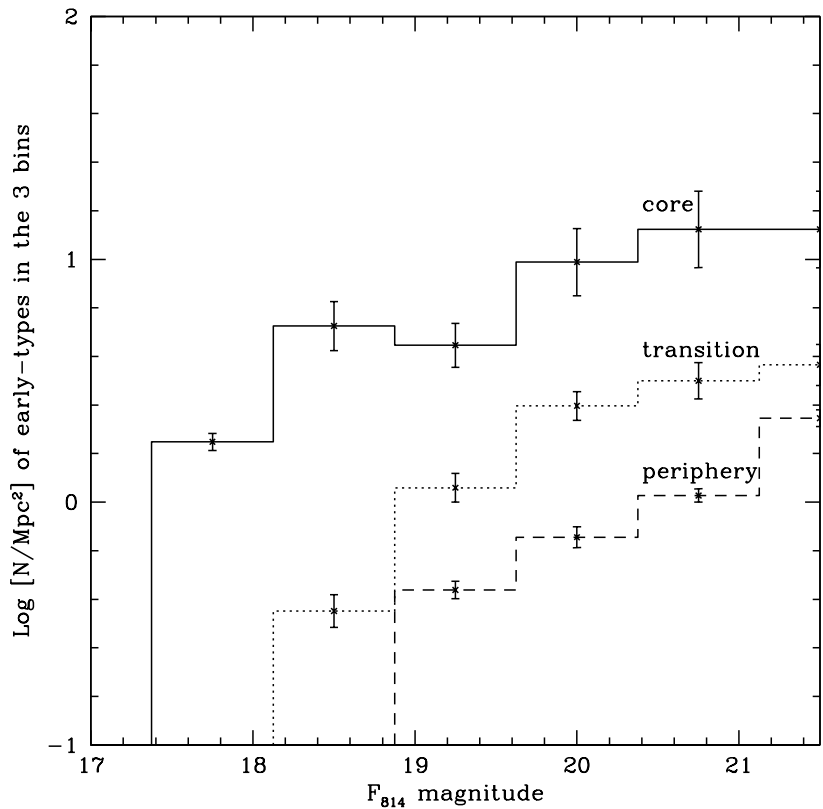,width=0.7\textwidth}}
\caption{The luminosity function of early-type galaxies in the 3
  regions: the number of galaxies per unit area versus magnitude is
  shown. It is clear from this plot that there is no systematic
  luminosity selection bias with cluster-centric radius for the
  early-type cluster members. However, luminosity segregation is
  evident in the core region. The luminosity function plotted above
  includes spectroscopically confirmed cluster members and those with
  secure photometric redshifts in the {\it HST} footprint.}
\end{figure*}

For each radial bin and type, we applied the likelihood analysis
described above to extract the best-fit parameters and significance
bounds for the dark matter halo associated with a fiducial $L^*$
subhalo in the cluster. Gravitational lensing effects are sensitive to
the total mass $M$ enclosed by a subhalo within an aperture $r_{t*}$. To account
for the differing mass-to-light ratios of the early and late-type
galaxies, we utilized the well-known empirical relations between the
velocity dispersion and luminosity for early-types (Faber-Jackson
relation); and equivalently that between the circular velocity and
luminosity (Tully-Fisher relation) for late-type galaxies. We used the
relations determined for Cl\,0024+16 by Moran et al. (2005; 2007) to relate
mass and light in our modeling procedure. An L$^*$ early-type galaxy
and late-type galaxy are assumed to have the same luminosity. The
limitations and systematics of the galaxy-galaxy lensing analysis in
clusters has been described in detail in our all our earlier papers,
below we briefly mention some of the key uncertainties of this method.

The following basic tests were performed for Cl\,0024+16, (i) choosing
random locations (instead of bright, cluster member locations) for the
perturbers; (ii) scrambling the shapes of background galaxies; and
(iii) choosing to associate the perturbers with the faintest (as
opposed to the brightest) galaxies. None of the above yields a
convergent likelihood map, in fact all that is seen in the resultant
2-dimensional likelihood surfaces is noise for all the above test
cases.

While the robustness of our method has been extensively tested and
reported in detail in earlier papers, there are a couple of caveats
and uncertainties inherent to the technique that ought to be
mentioned. In galaxy-galaxy lensing we are only sensitive to a
restricted mass range in terms of secure detection of
substructure. This is due to the fact that we are quantifying a
differential signal above the average tangential shear induced by the
smooth cluster component. Therefore, we are inherently limited by the
average number of distorted background galaxies that lie within the
aperture scale radii of cluster galaxies. This trade-off between
requiring a sufficient number of lensed background galaxies in the
vicinity of the subhalos and the optimum locations for the subhalos
leads us to choose the brightest cluster galaxies in each radial bin.
It is possible that the bulk of the mass in subhalos is in lower mass
clumps, which in this analysis is essentially accounted for as being
part of the smooth components. Also we cannot sensibly quantify the
contribution of close pairs/neighbors individually as it is essentially
a statistical technique.

Our results are robust and we statistically determine the mass of a
dark matter subhalo that hosts an $L^*$ galaxy. Even if we suppose
that the bulk of the dark matter is associated with very low
surface brightness galaxies in clusters, the spatial distribution of
these galaxies is required to be fine-tuned such that these effects do
not show up in the shear field in the any of the 3 regions. In
summary, the principal sources of uncertainty in the above analysis
are (i) shot noise -- we are inherently limited by the finite number
of sources sampled within a few tidal radii of each cluster galaxy;
(ii) the spread in the intrinsic ellipticity distribution of the
source population; (iii) observational errors arising from
uncertainties in the measurement of ellipticities from the images for
the faintest objects and (iv) contamination by foreground galaxies
mistaken as background.

The shot noise is clearly the most significant source of error,
accounting for up to $\sim 50$ per cent; followed by the width of the
intrinsic ellipticity distribution which contributes $\sim 20$ per
cent, and the other sources together contribute $\sim 30$ per cent
(Natarajan, De Lucia \& Springel 2007). This inventory of errors
suggests that the optimal future strategy for such analyses is to go
significantly deeper and wider in terms of the field of view.

\section{Results from galaxy-galaxy lensing}

The fiducial mass of a dark matter subhalo hosting an $L^*$ early-type
galaxy in the central region contained within an aperture of size
$r_{t*} = 45\pm 5\,{\rm kpc}$ is $M =
6.3_{-2.0}^{+2.7}\,\times\,10^{11}\,M_{\odot}$; in the transition
region it increases to $M =
1.3_{-0.6}^{+0.8}\,\times\,10^{12}\,M_{\odot}$, and in the periphery
it increases further to $M =
3.7_{-1.1}^{+1.4}\,\times\,10^{12}\,M_{\odot}$. All error bars
represent 3-$\sigma$ values. These values derived from the likelihood
analysis are shown in Figure~3. The increasing masses of the subhalos
with cluster radius demonstrates that the subhalos that host $L^*$ galaxies 
in the inner regions (core and transition) are subject to more severe tidal
truncation than those in the periphery. The mass of a typical subhalo
that hosts an $L^*$ early-type galaxy increases with cluster-centric
radius in concordance with theoretical expectations. The dark matter
subhalo associated with a typical late-type galaxy in the transition
and peripheral region is detected, with an aperture mass of $M =
1.06_{-0.41}^{+0.52}\,\times\,10^{12}\,M_{\odot}$ enclosed within a
radius of $r_{t*} = 25\,\pm 5 {\rm kpc}$ (shown as the solid triangle
in Figure~3). The total mass-to-light ratio for these fiducial subhalos
can also be estimated. The constant total mass-to-light ratio curves
over-plotted on Figure~3, suggest that a typical subhalo hosting an
L$^*$ early-type has a $M/L_{V} \sim 7, 10, 14$ respectively in the
three radial bins and a subhalo hosting an equivalent luminosity
late-type galaxy has a $M/L_V \sim 10$. These values suggest that
galaxies in clusters do possess individual dark matter subhalos that
extend to well beyond the stellar component.

Utilizing the scaling with luminosity provided by the Faber-Jackson
and Tully-Fisher relations, we derived the mass function of subhalos
within each bin (Figure~4). Clearly, the core region where the central
density of the cluster is maximal is expected to be an extreme and
violent environment for infalling galaxies. We interpret our results
to be a consequence of the fact that galaxies in the inner bin are
more tidally truncated as they likely formed earlier and have
therefore had time for many more crossings through the dense cluster
center.

\begin{figure*}
\centerline{\psfig{file=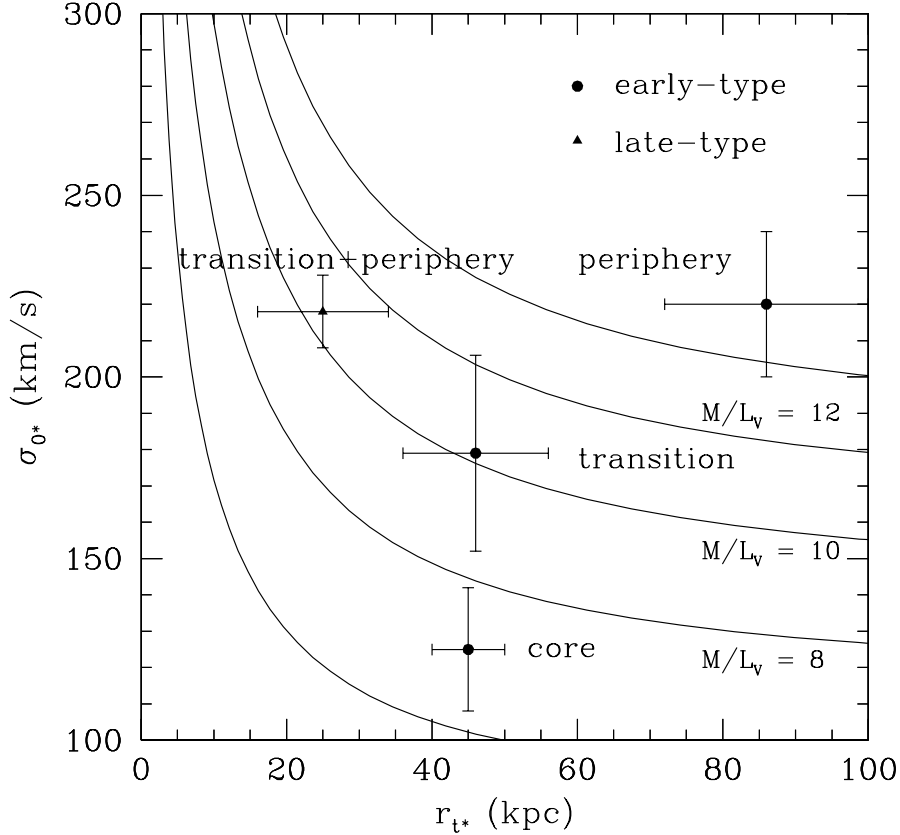,width=0.8\textwidth}}
\vspace{1cm}
\caption{The fiducial value of the central velocity dispersion
($\sigma_{0*}$) and aperture radius ($r_{t*})$ for an $L^*$
galaxy. These two parameters are chosen in the optimization for the
PIEMD fit to the subhalos. The mass of a subhalo in the context of
this model is proportional to $\sigma_{0*}^2\,r_{t*}$.  Over-plotted
are curves of constant mass-to-light ratio in the V-band with values
6, 8, 10, 12 and 14 (increasing from the bottom up). The solid circles
are for subhalos associated with early-type galaxies and the
solid triangle symbol is for the subhalo associated with late-type
galaxies. The plotted error bars are 3-$\sigma$ derived from the
likelihood contours.}
\end{figure*}

These results are in very good agreement with theoretical predictions
wherein galaxies in the inner region are expected to be violently
tidally stripped of their dark matter content, while those in the
periphery are unlikely to have had even a single passage through the
cluster center and therefore be untouched by tidal interactions. A
simple analytic model (Merritt 1985) is used to predict the mass
enclosed within the tidal radius\footnote{The aperture radius $r_t$
that we infer from the lensing analysis is a proxy for the tidal
radius of a dark matter subhalo.} as a function of cluster-centric
distance, and is found to be consistent with our results (solid line
in Figure~6). Our results are also consistent with the findings of Gao
et al.(2004), who found that subhalos closer to the cluster center
retain a smaller fraction of their dark matter. Furthermore, we are
able to quantify the dark matter subhalo masses associated with
late-type galaxies in Cl\,0024+16.

 While the mean mass of a dark matter subhalo associated with an
early-type cluster galaxy increases with cluster-centric distance out
to 5 Mpc and they trace the overall spatial distribution of the smooth
mass components robustly.  The subhalos associated with late-type
galaxies do not contribute significantly to the total subhalo mass
function at any radius. In fact, it appears that the host subhalos of
late-types do not trace the total dark matter distribution in
clusters. We infer that the mass within 5 Mpc in Cl\,0024+16 is
distributed as follows: $\sim$70\% of the total mass of the cluster is
smoothly distributed, the subhalos associated with early-type galaxies
contribute $\gsim\,20\%$, and subhalos hosting late-type galaxies
account for the remaining $<\,10\%$.

\subsection{Comparison with N-body simulations}

In this section, we compare the lensing results discussed above with
results from the Millennium Simulation (Springel et al. 2005). The
simulation follows $N = 2160^3$ particles in a box of size
$500\,h^{-1}\,{\rm Mpc}$ on a side, with a particle mass of
$8.6\times10^{8}\,h^{-1}{\rm M}_{\odot}$ (yielding several hundred
particles per subhalo), and with a spatial resolution of $5\,
h^{-1}\,{\rm kpc}$.  For each snapshot of the simulation (in total
$64$), substructures within dark matter halos have been identified
using the algorithm {\small SUBFIND} (Springel et al. 2001). We refer
to the original paper for more details on the algorithm. Further
details of the determination of subhalo masses and the biases therein
are discussed in Natarajan, De Lucia \& Springel (2007).

For our comparison with Cl\,0024+16, we have selected all cluster
halos with $M_{200} \ge 8\times10^{14}\,{\rm M}_{\odot}$ from the
simulation box at $z\sim0.4$. A total of $12$ such cluster scale halos
are found. We then use the publicly\footnote{A description of the
publicly available catalogues, and a link to the database can be found
at the following webpage: http://mpa-garching.mpg.de/millennium/}
available results from the semi-analytic model described in De Lucia
\& Blaizot (2007) to select all galaxies in boxes of $10\,h^{-1}\,{\rm
Mpc}$ on a side and centred on the selected halos. We note that the
following nomenclature is used for galaxies in the adopted
semi-analytic model: each FOF group hosts a `central galaxy' (Type 0)
that is located at the position of the most bound particle of the main
halo. All other galaxies attached to dark matter subhalos are labeled
as Type 1 and located at the positions of the most bound particle of
the parent dark matter substructure. Tidal truncation and stripping
can disrupt the substructure down to the resolution limit of the
simulation. A galaxy that is no longer identified with a dark matter
subhalo is labeled as Type 2, and it is assumed not to be affected by
processes that reduce the mass of its parent subhalo. The positions of
Type 2 galaxies are tracked using the position of the most bound
particle of the subhalo before it was disrupted.

We then select all galaxies brighter than $M_K = -18.3$ (this
corresponds to all galaxies brighter than $1/20*L^*$, as $M_{K*} =
-21.37$) and classify as early-types those with $\Delta M = M_B -
M_{bulge} < 0.4$, where $M_B$ is the B-band rest-frame magnitude and
$M_{bulge}$ is the B-band rest frame magnitude of the bulge (Simien \&
de Vaucouleurs 1986). For each simulated cluster, we consider the same
three radial bins used for our lensing analysis and stack the results
for the projections along the $x$, $y$, and $z$ axes. This is done to
mimic as best the projected distances that we employ in our lensing
analysis in the 3 radial bins. In addition, we only consider galaxies
within 1 Mpc (in the redshift direction) from the cluster centre along
the line-of-sight. This choice is motivated by the width of the
measured velocity dispersion histogram in Cl\,0024+16 and therefore
reduces contamination from unassociated structures. The inventory is
as follows: Core region -- the models predict a total of $\sim$ 74
early-types that make the selection cut of which 42 are Type 2
galaxies and 32 are Type 0 \& 1's; Transition region -- the models
predict a total of 83 early-type galaxies that make the selection cut
of which 41 are Type 2 galaxies and 42 are Type 0 \& 1's; Outer region
-- the models predict a total of 22 early-types that make the
selection cut of which 10 are Type 2 galaxies and 12 are Type 0 \&
1's. In contrast, the selection from the observational data of
Cl\,0024+16 yields the following numbers for spectroscopically
confirmed early-types with equivalent selection criteria: Core region
-- 51 early-types; Transition region -- 97 early-types; Outer region
-- 47 early-types.

In Figure~4, we plot the luminosity function of early type galaxies in
the three radial bins considered in this analysis from observations
(thick, solid histograms) and the model (thin solid and dashed
histograms). The thick solid histograms show the luminosity function
of the spectroscopically confirmed early-types in each bin. The thin,
solid histograms show the total model luminosity function for
equivalently selected early-types (this includes Type 1's, Type 0's
and Type 2 galaxies). The dashed histograms show the luminosity
function of Type 2 galaxies only. We note that in all regions the
contribution by number of Type 2 galaxies is comparable to that of
Type 0 and Type 1's. In the core region 58\% of all model early-types
are Type 2's, in the transition region 50\% of all model early-types
are Type 2's and in the outer region 46\% of all model early-types are
Type 2 galaxies.

\begin{figure*}
\centerline{\psfig{file=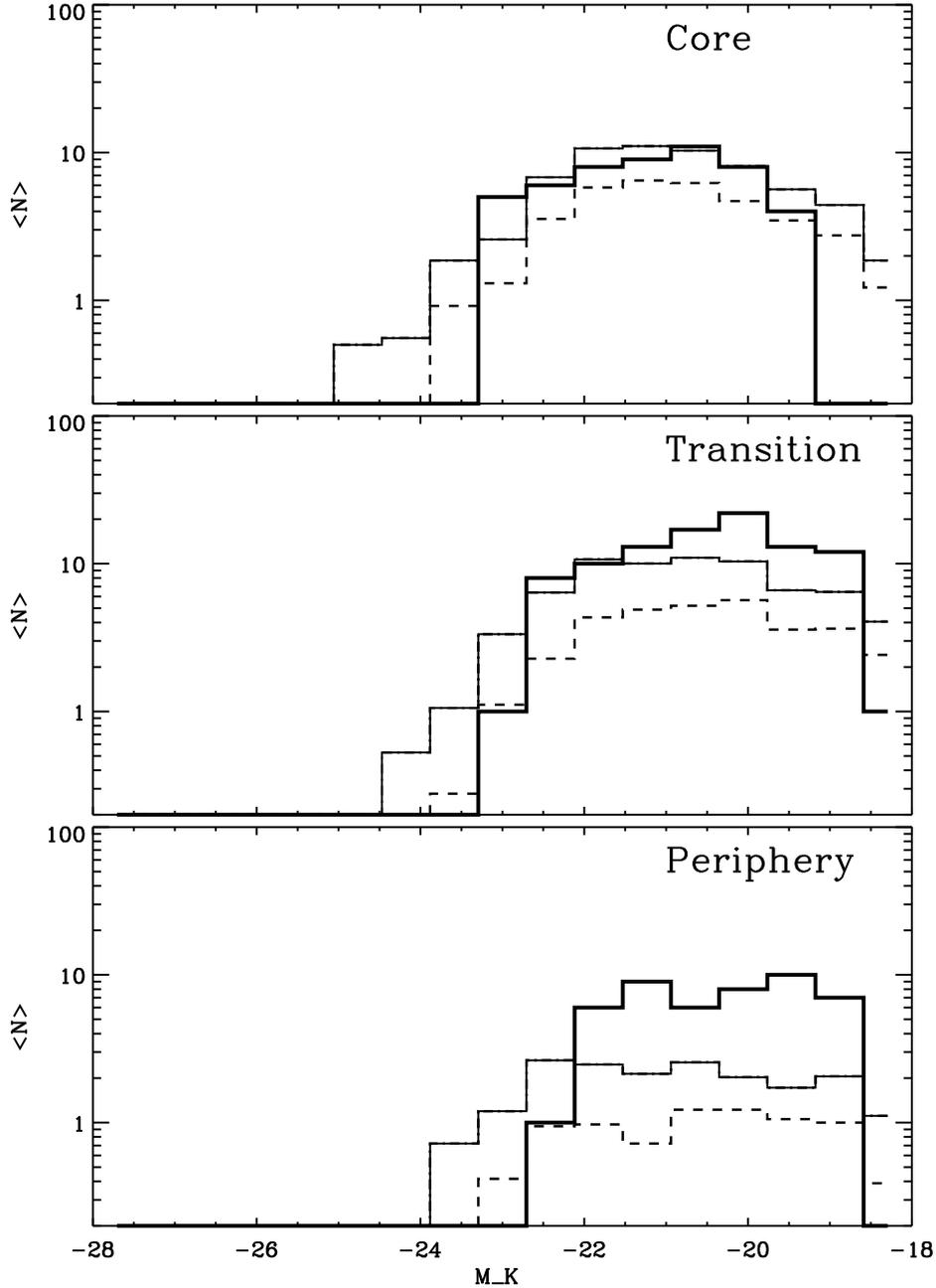,width=0.8\textwidth}}
\vspace{1cm}
\caption{The luminosity function of the early-type galaxies in the
  three radial bins considered in this analysis. The value of $M_{K*}$
  in the K-band is -21.37. Note that the y-axis for the model galaxies
  is an averaged number $<N>$ as the total number of selected
  early-types is divided by 36 to take into account 12 clusters each
  with 3 independent projections. For the observed galaxies the y-axis
  denotes the raw number. In all panels, the thick solid histograms
  show the luminosity function of spectroscopically confirmed
  early-type members from the Cl0024+16 data set. The thin solid
  histograms are total luminosity function of early-types in the
  simulations including Type 0, Type 1 and Type 2 galaxies. We
  separately show the luminosity function of Type 2 early-type
  galaxies as the dashed histograms. The fraction of Type 2 galaxies
  in the core is 58\%, in the transition region it is 50\% and in the
  outer regions it is 46\%.}
\end{figure*}

It is clear from Figure~4 that the luminosity functions of the
early-type galaxies in simulations agree rather well with the observed
ones in all 3 bins. However, we note that the inability of the lensing
analysis to accomodate/distinguish Type 2 galaxies which constitute
roughly half the number of early-types in the model will limit our
analysis. This discrepancy in the total number of early-types that are
hosted in individual dark matter halos in the model versus the lensing
analysis will re-appear when we compare the masses of a typical
subhalo that hosts an $L^*$ early-type galaxy.

In Figure~5 we compare the mass function of dark matter subhalos
obtained from the galaxy-galaxy lensing analysis with that obtained
from averaging the 12 massive clusters (each with 3 independent
projections) in the Millennium Simulation. In the left hand panel, we
plot a direct comparison of the mass functions without taking into
account the discrepancy in number between the observations and the
simulations. In the right hand panel, we scale the observations to
take into account the fraction of Type 2's versus Type 0 \& 1's in
each radial bin. This is done by normalizing the observations to match
the fraction of Type 0s and Type 1s.In order to make a sensible
comparison, we focus on the right hand panel of Figure~5. In the core
region, the mass function from simulations agrees quite nicely with
that determined using the galaxy-galaxy lensing analysis. It is
notable that general shapes of the mass function are in very good
agreement. The agreement between the shape of the mass functions in
transition and outer regions is also good.\footnote{It is worth
mentioning here that in Natarajan, De Lucia \& Springel (2007)
Cl\,0024+16 was the one outlier from the general good agreement. The
lensing determined mass function for Cl\,0024+16 within 1 Mpc was not
in good agreement with the subhalo mass function derived from
simulations. We attribute the current agreement of the mass functions
in the core region ($r < 0.6 Mpc$) to the following 2 key factors (i)
careful selection based on early-type cluster members to mimic the
observations and (ii) more careful classification into Type 1 and Type
2 galaxies and taking their relative numbers into account when
computing the mass function.}

In the inner region, we found in earlier work (Natarajan, De Lucia \&
Springel 2007) that the masses from the simulation tend to be
under-estimated by a factor 2 or so.  This offset was found in our
earlier analysis of the core regions in 5 clusters (r < 1 Mpc)
reported in Natarajan, De Lucia \& Springel (2007). The origin of this
offset in the core region has to do primarily with the systematics due
the method employed to determine subhalo masses. The SUBFIND algorithm
used to find dark matter substructures tends to underestimate their
masses (for a more extensive discussion and diagnostic plots see
Figure~3 in Natarajan, De Lucia \& Springel 2007) by a factor of 2 in
the inner regions. Natarajan, De Lucia \& Springel (2007) have
also shown that the cluster-to-cluster variation for simulated halos
is quite large. Therefore, we do not correct for this bias in the current
analysis of Cl\,0024+16.

We note here that the subhalo mass function available from the
simulations and the lensing technique probe a comparable mass range
suggesting that our early-type galaxy selections have been
equivalent. In the periphery, subhalo masses derived from lensing are
of the order of few times $10^{13}\,\msun$, which is typical of group
masses, suggestive of the presence of infalling groups. Tracking
morphological types and their transformations in this region Treu et
al. (2003) also suggest the prevalence of infalling groups, in
consonance with our lensing results.So we emphasize here, that in the
outskirts of the Cl\,0024+16, it appears that inferred subhalo masses
correspond to group scale masses suggesting that these halos likely
contain other fainter galaxies in addition to the bright, early-type
that we tag in this analysis.

\begin{figure*}
\begin{minipage}{3.5in}
\centerline{\psfig{file=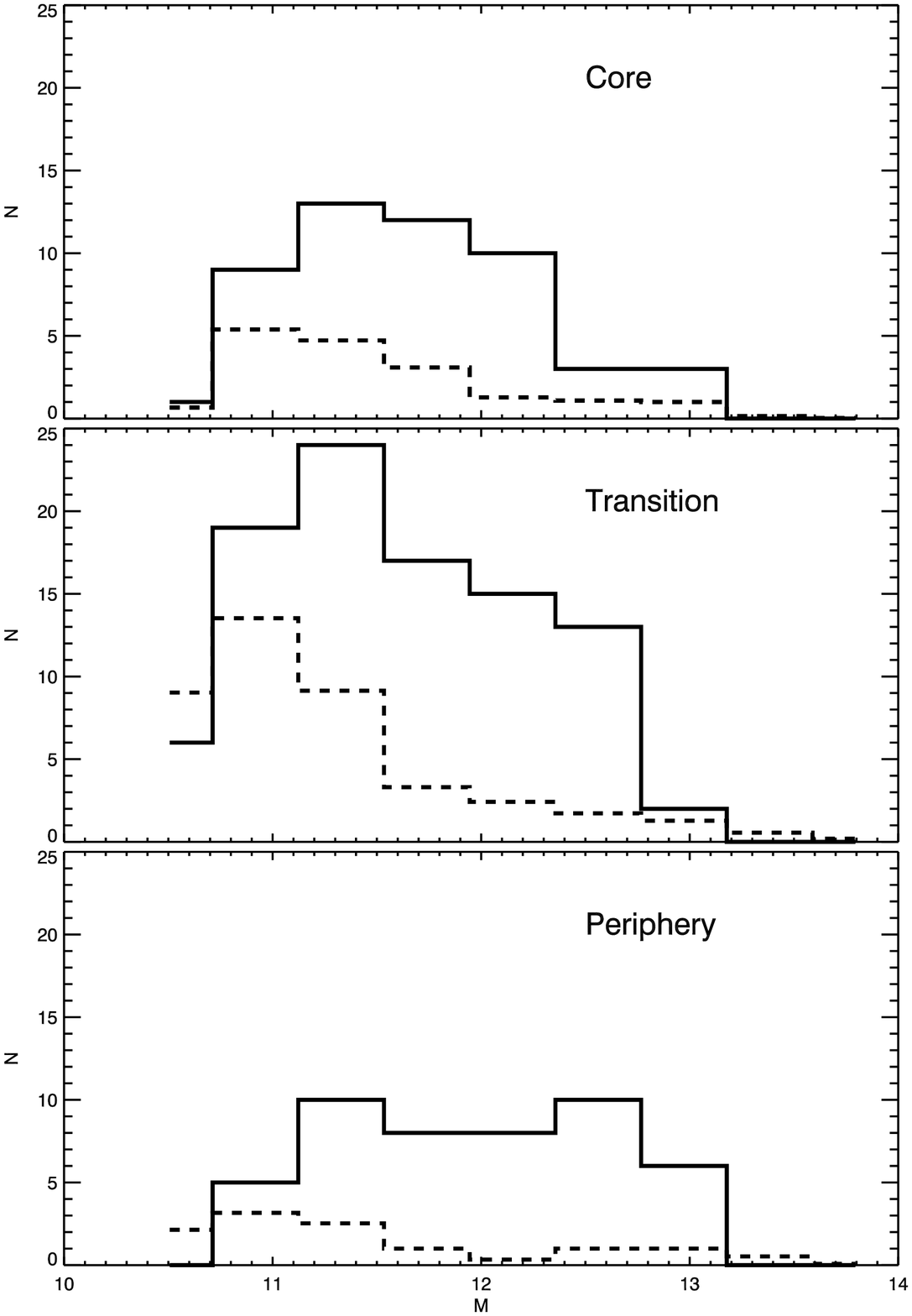,width=3.5in}}
\end{minipage} \hspace*{0in}\begin{minipage}{3.5in}
\psfig{file=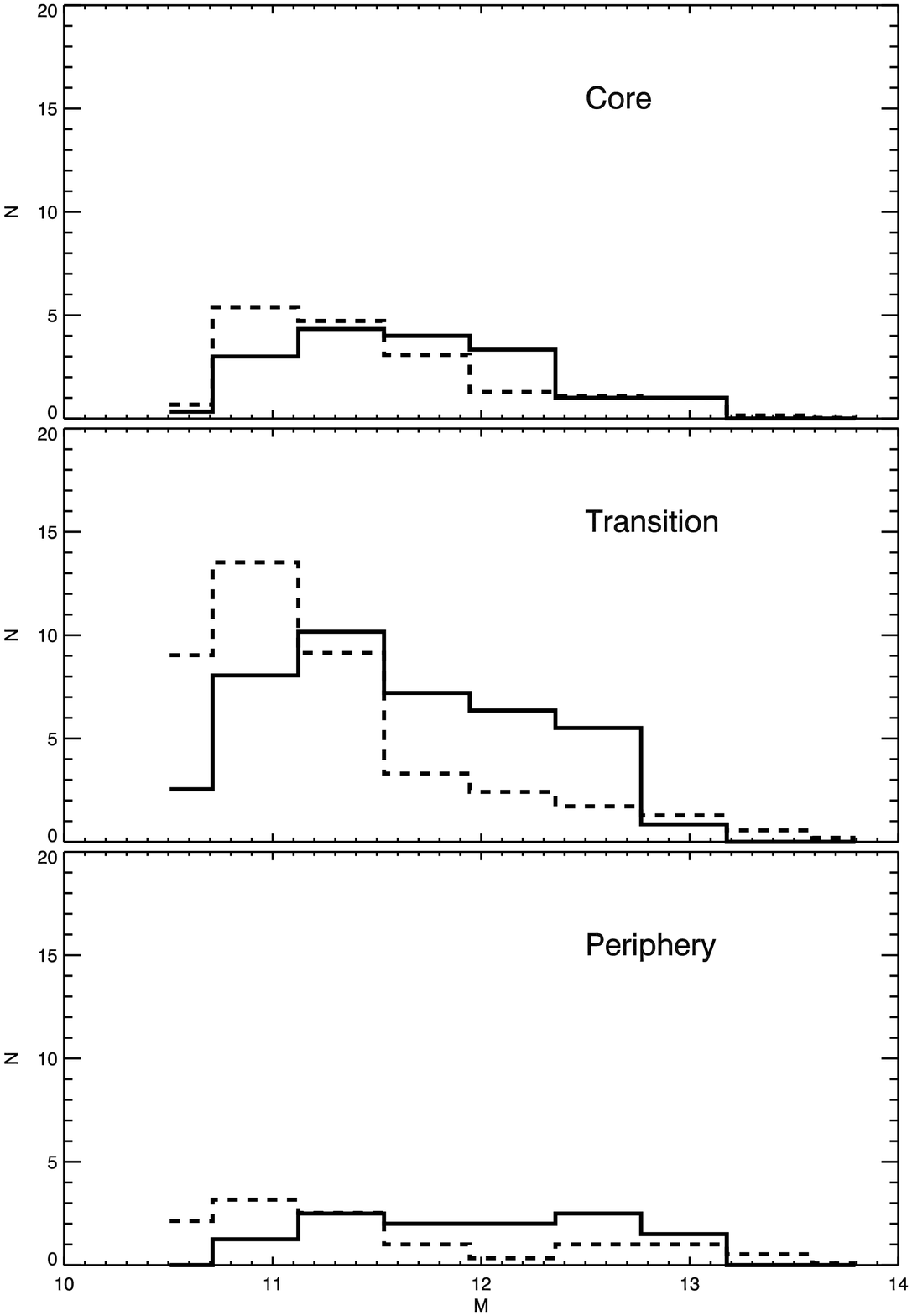,width=3.5in}
\end{minipage}
\caption{Comparison of the mass function determined from galaxy-galaxy
lensing as a function of cluster-centric distance with that determined
from simulated clusters in the Millenium simulation. The solid
histograms are results from the lensing analysis and the dashed ones
are from the Millenium Run. The raw mass function without any
normalization or scaling is shown in the right hand panel, whereas in
the left hand panel the lensing derived mass function is normalized to
compare with the model mass functions.}
\end{figure*}

In Figure~6, we plot the mass of a typical subhalo that hosts an
early-type L$^*$ galaxy as a function of cluster-centric radius
derived from galaxy-galaxy lensing and the simulations. The solid line
in Figure~6 is the trend derived from a simple analytic model of tidal
stripping of galaxies by an isothermal cluster proposed by Merritt
(1985). The dashed-line is the also the analytic model offset
appropriately to compare with the simulation results. The model curve
is nearly identical in slope to the best-fit line going through the
simulation points. The radial trends are in very good agreement
although there is an offset of a factor of $\sim 2.5$ in the mean
value of the subhalo masses. It is likely that systematics in the
lensing also contribute to this discrepancy. The efficiency of tidal
stripping depends on the central density of the cluster. Writing this
out explicitly, we have:$$ M_{\rm lens}/M_{\rm sim} \sim 2.5$$. The
offset in subhalo masses by a factor of $\sim 2.5$ suggests that the
tidal stripping in the averaged simulated clusters is more efficient
than in Cl\,0024+16 as inferred from the lensing data. We note here
that the values of $M_{200}$ for the 12 simulated clusters range from
$\sim\,8\,\times\,10^{14}\,\msun$ to $\sim\,2\,\times\,
10^{15}\,\msun$ and the best-fit parameters for Cl\,0024+16 from
observations which consist of a super-position of 2 NFW profiles with
$M_{200} \sim 4 \times 10^{14} \msun$ and $M_{200} \sim 1.8 \times
10^{14} \msun$, could partially account for the discrepancy. The
simulated ensemble does not reproduce the observed bi-modal mass
distribution in Cl\,0024+16 which has important dynamical
consequences. No dynamical analog to Cl\,0024+16 was found in the
Millenium Run at $z \sim 0.4$. Therefore it is not surprising that
there is a discrepancy in the inferred mass for a dark matter subhalo
hosting an $L^*$ galaxy. Note that if we correct the subhalo mass in
the inner most bin by a factor of 2 as found in our earlier work, the
agreement gets significantly better in the core region consistent with
our earlier results (Natarajan, De Lucia \& Springel
2007). Regardless, there appears to be an offset despite overall
agreement in the ensemble mass functions (as shown in the right hand
panel of Figure~5). Tidal stripping of dark matter appears to be
more efficient in the simulations compared to estimates from the
lensing data.

We note here that in the Millenium simulation only the dark matter is
followed dynamically but not the baryons. It has been recently argued
that the adiabatic contraction of baryons in the inner regions of
galaxies and clusters is likely to modify density profiles
appreciably. Such modifications will impact the efficiency of tidal
stripping in clusters. This claim has been made in numerical
simulations that include gas cooling and prescriptions for star
formation by Gnedin et al. (2004). Zappacosta et al. (2006) on the
other hand claim using the case of the cluster Abell 2589 that
adiabatic contraction is unimportant for the overall mass distribution
of clusters. A recent study by Limousin et al.(2007b) that examines
the tidal stripping of subhalos in numerical simulations and includes
baryons, find a radial trend in the mass function that is in good
agreement with the lensing derived trend in Cl\,0024+16.

Meanwhile in lensing the systematic arises from the fact that we do
not have measured redshifts for all background sources. While the mass
calibration is most sensitive to the median redshift adopted for the
background galaxies, biases are introduced if the median redshift is
over-estimated or under-estimated. 

\begin{figure*}
\centerline{\psfig{file=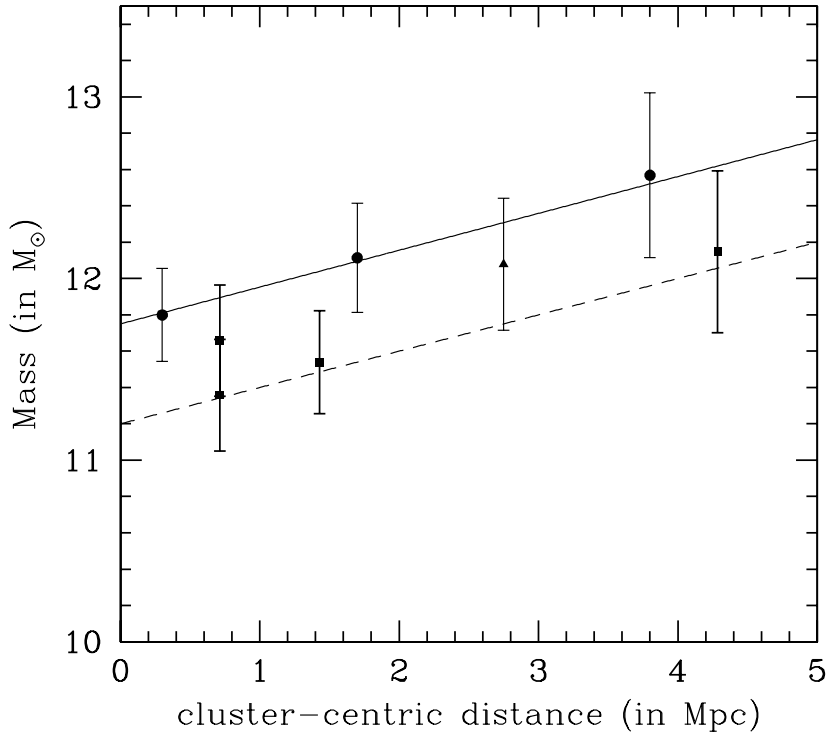,width=0.8\textwidth}}
\caption{The variation of the mass of a dark matter subhalo that hosts
an early-type L$^*$ galaxy as a function of cluster centric
radius. The results from the likelihood analysis are used to derive
the subhalo mass for the galaxy-galaxy lensing results and the
counterparts are derived from the Millenium simulation with an
embedded semi-analytic galaxy formation model. This enables selection
of dark matter halos that host a single L$^*$ galaxy akin to our
assumption in the lensing analysis. The solid circles are the data
points from the galaxy-galaxy lensing analysis and the solid squares
are from the Millenium simulation. The upper solid square in the core
region marks the value of the subhalo mass with correction by a factor
of 2 as found in Natarajan, De Lucia \& Natarajan (2007). The solid
triangle is the galaxy-galaxy lensing data point for the subhalo
associated with a late-type $L^*$ galaxy. The radial trend derived
from lensing is in very good agreement with simulations although there
is an offset in the masses which is discussed further in the text.}
\end{figure*}

We note that the galaxy-galaxy lensing technique is sensitive to the
detection of subhalo masses above a threshold value that is determined
by the quality of the observational lensing data. The selection made
in the cluster luminosity function translates into a mass limit. The
contribution of fainter early-types (galaxies fainter than our
selection limit) translates into lower mass subhalos due to the
assumed luminosity scalings. As a consequence, subhalos with lower
masses get included in the mass inventory as constituting the `smooth'
component. The lensing derived mass functions are therefore complete
at the high mass end but are typically incomplete at the low mass
end. The cut-off at the low mass end is hence determined primarily by
the depth of the observational data and the ability to measure shapes
accurately for the faintest background sources. Since galaxy-galaxy
lensing analysis in clusters is inherently statistical, its robustness
is also limited by the ability to accurately pin down the smooth mass
component, subtract it from the observed shear field and then stack
the residuals to characterize the mass of a detectable dark matter
subhalo. In the inner region while the constraints on the smooth
component are tighter due to the presence of strong lensing features,
the smooth component also tends to dominate the overall mass
distribution, so subtracting it is challenging. In the outer regions
while the smooth component is sub-dominant, there are fewer
constraints and the overall value of the shear is significantly lower
as well. These trade-offs cause a varying mass resolution for the
lensing technique as a function of cluster-centric radius. However,
since we assume scaling relations with luminosity, and use the cluster
galaxy luminosity function to determine the mass function, the
detectable limit of subhalo masses is set predominantly by the
magnitude cut adopted for the selected early-type galaxies. Note that
in our comparison with simulations we have restricted ourselves only
to early-types with measured spectroscopic redshifts. Since
spectroscopic follow-up tends to be easier for brighter galaxies, once
again our lensing derived mass function is more complete at the high
mass end and is less so at the low mass end. However, from the
comparison of the mass functions we note that both methods lensing and
the simulations are probing comparable subhalo mass ranges.

\section{Discussion and Conclusions}

Earlier work on galaxy-galaxy lensing in the field has identified a
signal associated with massive halos around typical field galaxies,
extending to beyond 100\,kpc (e.g.\ Brainerd, Blandford \& Smail 1996;
Ebbels et al.\ 2000; Hudson et al.\ 1998; Wilson et al.\ 2001;
Hoekstra et al.\ 2004). In particular, Hoekstra et al.\ (2004) report
the detection of finite truncation radii via weak lensing by galaxies
based on 45.5 deg$^2$ of imaging data from the Red-Sequence Cluster
Survey. Using a truncated isothermal sphere to model the mass in
galaxy halos, they find a best-fit central velocity dispersion for an
$L^*$ galaxy of $\sigma = 136 \pm 5$ kms$^{-1}$ (68\% confidence
limits) and a truncation radius of $185 \pm 30$ kpc. Galaxy-galaxy
lensing results from the analysis of the Sloan Digital Sky Survey data
(Sheldon et al. 2004; Guzik \& Seljak 2002; Mandelbaum et al. 2006)
have contributed to a deeper understanding of the relation between
mass and light. Similar analysis of galaxies in the cores of rich
clusters suggests that the average mass-to-light ratio and spatial
extents of the dark matter halos associated with
morphologically-classified early-type galaxies in these regions may
differ from those of comparable luminosity field galaxies (Natarajan
et al.\ 1998, 2002a). We find that at a given luminosity, galaxies in
clusters have more compact halo sizes and lower masses (by a factor of
2--5) compared to their field counter-parts. The mass-to-light ratios
inferred for cluster galaxies in the V-band are also lower than that
of comparable luminosity field galaxies. This is a strong indication
of the tidal stripping effect of the dense environment on the
properties of dark matter halos. In recent work, using only strong
lensing constraints in the inner regions of the Abell cluster A\,1689
derived from images taken by the Advanced Camera for Surveys (ACS)
aboard {\it HST}, Halkola \& Seitz (2007) also find independently that
the subhalos of cluster galaxies are severely truncated compared to
equivalent luminosity galaxies in the field.

 The subhalo mass function represents an important prediction of
 hierarchical CDM structure formation models and has been subject of
 intense scrutiny since the `satellite crisis' was identified (Moore
 et al.\ 1999; Klypin et al.\ 1999). This crisis refers to the fact
 that within a radius of $400\,h^{-1}\,{\rm kpc}$, from the Milky Way,
 cosmological simulations of structure formation predict $\sim$ 50
 dark matter satellites with circular velocities in excess of
 $50\,{\rm kms^{-1}}$ and mass greater than
 $3\,\times\,10^8\,M_{\odot}$.  This number is significantly higher
 than the dozen or so satellites actually detected around our
 Galaxy. Several explanations have been proposed to resolve this
 discrepancy.  The missing satellites could for example be identified
 with the detected High Velocity Clouds (Kerr \& Sullivan 1969;
 Willman et al. 2002; Maller \& Bullock 2004). {\it Warm} or {\it
 self-interacting} dark matter could also selectively suppress power
 on the small scales, therefore reducing the predicted number of
 satellites.  The leading hypothesis however remains that the solution
 to this problem lies in processes such as heating by a photo-ionizing
 background that preferentially suppresses star formation in small
 halos at early times (Bullock et al.\ 2000; Benson et al.\ 2002;
 Kravtsov et al. 2004). On the scale of galaxy clusters, many more
 dark matter structures are expected to be visible, thus making the
 comparison with expectation from numerical simulations less affected
 by uncertainties in the poorly understood physics of the galaxy
 formation. In earlier work, we showed that there is very good
 agreement between the lensing observations and the Millenium Run
 simulations in the inner 1 Mpc or so for a sample of clusters
 (Natarajan, De Lucia \& Springel 2007). Now we are able to extend our
 analysis out to 5 Mpc for Cl\,0024+16 due to the unique data-set that
 is available. The diagnostic available for comparison here is the
 shape of the mass function in each of the three bins. We find that
 the overall shape of the subhalo mass functions derived from the 2
 independent methods is in very good agreement out to beyond the
 virial radius. We find that the mass of a typical subhalo that hosts
 an $L^*$ early-type galaxy increases with cluster-centric radius in
 concordance with theoretical expectations. However, the estimates of
 the mass of a subhalo that hosts an $L^*$ galaxy derived from
 simulations is significantly lower than those derived from lensing
 observations. The origin of this discrepancy lies in the fact that
 tidal stripping appears to be more efficient in the simulations.

Due to the large area probed by this Cl\,0024+16 dataset, we are also
able to constrain the properties of dark matter subhalos associated
with late-type galaxies that preferentially lie in the outer regions
of the cluster (Treu et al. 2003). We report the first detection of
the presence of a dark matter subhalo associated with late-type
galaxies in Cl\,0024+16. While early-type galaxies appear to trace the
overall mass distribution robustly, the subhalos associated with
late-type galaxies do not contribute significantly to the total mass
budget at any radius. In the cluster Cl\,0024+16 within 5 Mpc we find
the following contributions to the total mass: $\sim$70\% of the total
mass of the cluster is smoothly distributed, the subhalos associated
with early-type galaxies contribute $\gsim\,20\%$, and subhalos
hosting late-type galaxies account for the remaining $<\,10\%$.

The mass resolution of our technique varies slightly with
cluster-centric distance owing to the nature of observational
constraints that dominate the likelihood optimization. While the
strong lensing constraints in the core are the most stringent and
drive the fit in the inner regions, the anisotropy in the shear field
is statistically harder to recover. As we progressively step out in
radius away from the cluster center, the shear of the large scale
smooth component drops, and that of the individual subhalos dominates
but the overall summed shear signal is significantly lower than in the
inner regions. The current analysis is primarily limited by the
quality of the available data. Datasets from the {\it ACS} will allow
mass modeling of lensing clusters at even higher resolutions providing
increasing accuracy enabling better mapping of the lower mass end of
the mass functions of substructure. However, as described above a vast
complement of ground-based observations are also needed for this kind
of comprehensive analysis which is extremely time consuming. Ground
based data provided many important constraints, for instance, the
large number of measured central velocity dispersions for cluster
galaxies (Moran et al. 2007) were used as priors in modeling the
perturbing subhalos that made the optimization more efficient.

Below we summarize the key results on comparison with simulations,
where we mimic-ed the selection process adopted for the observational
data of Cl\,0024+16.  Dividing the simulated clusters drawn from the
Millenium Run into 3 equivalent radial bins as the observational data,
we were able to estimate (i) the mass function in each bin and (ii)
the subhalo masses that host $L^*$ early-type galaxies. The shapes of
the lensing derived mass functions are in reasonable agreement
with those derived from simulations when we normalize the lensing
results to the those of the total number of model Type 1's and Type
2's..

Our results provide strong support for the tidal stripping
hypothesis. We also find evidence for the variation in the efficiency
of tidal stripping with cluster-centric radius and morphological type.
We conclude that dark matter in clusters is assembled by the
incorporation of infalling subhalos that are progressively stripped
during their journey through the cluster. The finding of
kinematically disturbed features in the cluster galaxy population by
Moran et al. (2007) corroborates our conclusion. We have significantly
improved on previous ground-based studies as space-based data affords
greater accuracy in shape measurements. Future space-based surveys
coupled with ground-based spectroscopic follow-up will provide an
unprecedented opportunity to follow the cluster assembly process.

\medskip

Acknowledgments: Pranjal Trivedi is thanked for help during the early
stages of this project. Gabriella De Lucia is gratefully acknowledged
for her help in extracting catalogs from the Millenium Simulation, in
the analysis of simulated data and thoughtful comments on the
draft. Help from Volker Springel is acknowledged for clarifications
regarding sub-halo extraction from the Millenium Simulation.  Some of
this work was conducted using support from the grant HST-GO-09722.06-A
provided by NASA through a grant from Space Telescope Science
Institute, which is operated, by the Association of Universities for
Research in Astronomy, Incorporated, under NASA contract
NAS5-26555. This paper is based on observations made with the NASA/ESA
Hubble Space Telescope, which is operated by the Association of
Universities for Research in Astronomy, Inc., under NASA contract NAS
5-26555.  IRS acknowledges support from the Royal Society. ML
acknowledges support from the Dark Cosmology Centre funded by the
Danish National Research Foundation. The Millennium Simulation
databases used in this paper and the web application providing online
access to them were constructed as part of the activities of the
German Astrophysical Virtual Observatory.

\section*{Appendix}

\subsubsection*{The intrinsic shape distribution of background galaxies}

As in all lensing work, it is assumed here as well, that the intrinsic or
undistorted distribution of shapes of galaxies is a known quantity. This
distribution is obtained from shape measurements taken from deep
images of blank field surveys. Previous analysis of deep survey
data such as the MDS fields (Griffiths et al.\ 1994) showed that the 
ellipticity distribution of sources is a strong function of the sizes 
of individual galaxies as well as their magnitude (Kneib et al.\
1996). For the purposes of our modeling, the intrinsic ellipticities
for background galaxies are assigned in concordance with an
ellipticity distribution $p_(\tau_S)$ where the shape parameter 
$\tau$ is defined as $\tau = (a^2-b^2)/(2ab)$ derived from the observed
ellipticities of the CFHT12k data (see Limousin et al.\ 2004; 2006; 
2007a for details): 
\be
p(\tau_S)\,=\,\tau_S\,\,\exp(-({\tau_S \over
\delta})^{\nu});\,\,\,\nu\,=\,1.15,\,\,\delta\,=\,0.25.  
\ee
This distribution includes accurately measured shapes of
galaxies of all morphological types. In the likelihood analysis
this distribution $p(\tau_S)$ is the assumed prior, which is used to
compare with the observed shapes once the effects of the assumed 
mass model are removed from the background images. We note here that 
the exact shape of the ellipticity distribution, i.e. the functional 
form and the value of $\delta$ and $\nu$ do not change the results, 
but alter the confidence levels we obtain. The width of the intrinsic
ellipticity distribution, on the other hand is the fundamental
limiting factor in the accuracy of all lensing measurements including
this work.

\subsubsection*{The redshift distribution of background galaxies}

While the shapes of lensed background galaxies can be measured
directly and reliably by extracting the second moment of the light
distribution, in general, the precise redshift for each weakly object
is in fact unknown and therefore needs to be assumed. Using
multi-waveband data from surveys such as COMBO-17 (Wolf et al.\ 2004)
photometric redshift estimates can be obtained for every background
object. Typically the redshift distribution of background galaxies is
modeled as a function of observed magnitude $P(z,m)$. We have used
data from the high-redshift survey VIMOS VLT Deep Survey (Le Fevre et
al.  2004) as well as recent CFHT12k R-band data to define the number
counts of galaxies, and the HDF prescription for the mean redshift per
magnitude bin, and find that the simple parameterization of the
redshift distribution used by Brainerd, Blandford \& Smail (1996)
still provides a good description to the data. We also used CFHT
K-band photometry to derive photometric redshifts and a $N(z)$ for all
the background sources in Cl\,0024+16 (Kneib et al. 2003; Smith et
al. 2005) but found that drawing instead from the distribution below
allowed us to go deeper to the magnitude limits required. We used the
derived photometric redshifts for sources with reliable estimates and
for the rest we drew from the distribution below. For the normalized
redshift distribution at a given magnitude $m$ (in the $I_{814}$ band)
we have: 
\bea 
N(z)|_{m}\,=\,{{\beta\,({{z^2} \over {z_0^2}})\,
\exp(-({z \over {z_0}})^{\beta})} \over {\Gamma({3 \over
\beta})\,{{z_0}}}}; 
\eea where $\beta\,=\,$1.5 and 
\bea
z_0\,=\,0.7\,[\,{z_{\rm median}}\,+\,{{d{z_{\rm median}}} \over
{dm_R}}{(m_R\,-\,{m_{R0}})}\,], 
\eea 
${z_{\rm median}}$ being the median redshift, $dz_{\rm median}/ dm_R$
the change in median redshift with say the $R$-band magnitude, $m_R$.

However, we note here in agreement with another study of
galaxy-galaxy lensing in the field by Kleinheinrich et al. (2004),
that the final results on the aperture mass presented here are also
sensitive primarily only to the choice of the median redshift of the
distribution rather than the individual assigned values. Since the
median redshift of the distribution we adopt here is similar to that
of COSMOS survey, our results would be robust.

\end{document}